\documentstyle[12pt,aasms,psfig]{article}

\def \etal {et al.\thinspace}

\received{ 2000}
\journalid{337}{15 January 1989}
\articleid{11}{14}

\slugcomment{Submitted to Astrophys. J. {\it Suppl.}, 2000}

\begin{document}

\title{Electron-Ion Recombination Rate Coefficients and Photoionization 
Cross Sections for Astrophysically Abundant 
Elements. V. Relativistic calculations for Fe~XXIV and Fe~XXV for X-ray
modeling}

\author{Sultana N. Nahar, Anil K. Pradhan}
\affil{Department of Astronomy, The Ohio State University, 
Columbus, OH 43210}
\author{Hong Lin Zhang}
\affil{Applied Theoretical \& Computational Physics Division}
\affil{MS F663, Los Alamos National Laboratory, Los Alamos, NM 87545}

\begin{abstract}

Photoionization and recombination cross sections and rate coefficients
are calculated for Li-like Fe~XXIV and He-like Fe~XXV using the 
Breit-Pauli R-matrix (BPRM) method. A complete set of total and 
level-specific parameters is obtained to enable X-ray photoionization 
and spectral modeling. The ab initio calculations for the unified (e~+~ion)
recombination rate coefficients include both the non-resonant and the
resonant recombination (radiative and di-electronic recombination, RR
and DR, respectively) for $(e + Fe~XXV) \longrightarrow Fe~XXIV$ and 
$(e + Fe~XXVI) \longrightarrow Fe~XXV$. The level specific rates are 
computed for all fine structure levels up to n = 10, enabling accurate 
computation of recombination-cascade matrices and effective rates for 
the X-ray lines.  The total recombination rate coefficients
for both Fe~XXIV and Fe~XXV differ considerably from the sum of RR and 
DR rates currently used to compute ionization fractions in 
astrophysical models.  As the photoionization/recombination calculations 
are carried out using an identical eigenfunction expansion, the cross 
sections for both processes are theoretically self-consistent; the 
overall uncertainty is estimated to be about 10-20\%. All data for 
Fe~XXIV and Fe~XXV (and also for H-like Fe~XXVI, included for 
completeness) are available electronically.

\end{abstract}


\section{INTRODUCTION}

Photoionization and recombination of Li-like and He-like ions is of 
particular interest in X-ray astronomy (e.g. Proceedings of the 
NASA workshop on 
{\it Atomic data needs in X-ray astronomy} 2000).  X-ray Photoionization
models in general require these parameters over all ranges of
photon energies and temperatures prevalent in high-temperature sources
such as the active galactic nuclei, supernova remnants, hot stellar
coronae (Canizares \etal 2000, Brickhouse and Drake 2000, Kaastra and Mewe
2000). High precision is
increasingly a prime requirement owing to the detailed observational 
spectroscopy from space obeservatories, particularly ASCA, 
CXO and XMM-Newton.
X-ray emission in the 6.7 keV K$\alpha$ complex of 
He-like Fe~XXV, from the $n = 2 \rightarrow 1$ transitions yields 
valuable spectral diagnostics for temperature, density, ionization 
balance, and abundances in the plasma source (Gabriel 1972, Mewe and 
Schrijver 1978, Pradhan and Shull 1981, Bely-Dubau \etal 1982, 
Pradhan 1985). Dielectronic satellite lines of Fe~XXIV 
formed due to (e~+~ion) recombination with Fe~XXV are useful as 
temperature diagnostics in both laboratory and astrophysical sources. 
Ionization balance and spectral studies therefore require comprehensive 
sets of photoionization/recombination cross sections.

For highly charged ions it is important to consider relativistic fine
structure explicitly in the theoretical formulation, in addition to the
often strong electron correlation effects. The close coupling
approximation, employing the R-matrix method, has been widely employed
to compute radiative and collisional parameters, such as under the Iron
Project (Hummer \etal 1993). 
The close coupling calculations for the He- and 
Li-like ionization states involve eigenfunction expansions for 
the H- and He-like target ions, respectively, and it is necessary to
include relativistic effects for high precision. Therefore, in order to improve 
the accuracy of available photoionization/recombination parameters,
as well as to provide extensive sets of data needed for astrophysical
models, we have re-calculated these using relativistic Breit-Pauli
R-matrix method (hereafter BPRM). The first such calculations were
reported for He- and Li-like C~V and C~IV in the present series (Nahar
\etal 2000), together with a detailed description of the relativistic
close coupling calculations for photoionization and recombination cross
sections (total and level-specific), and unified (e~+~ion) recombination 
rates. The present work is basically similar and extends the treatment 
to iron ions.

\section{THEORY}

The electron-ion recombination calculations entail close coupling 
calculations for photoionization and electron-ion scattering. Identical
eigenfunction expansion for the target (core) ion is employed for
both processes; thus enabling inherently self-consistent
photoionization/recombination results in an ab initio manner for a
given ion. General details of the theory and close coupling 
BPRM calculations for photoionization and recombination are 
described in Nahar \etal (2000), and references therein. We sketch below
the basic outline of the theoretical formulation.

We consider photoionization from, and recombination into, the 
infinity of levels of the (e~+~ion) system. These are divided into 
two groups of bound levels: (A) with $\nu \leq \nu_o$ and  all
possible fine structure SLJ symmetries, and (B) 
$ \nu_o < \nu \leq \infty $; where $\nu$
is the effective quantum number relative to the target threshold(s).
Photoionization and recombination calculations are carried out in 
detail for all group A levels. The photo-recombination cross sections 
are computed from the photoionization cross sections at a 
sufficiently large number of energies to delineate the non-resonant 
background  and the autoionizing resonances, thereby representing both 
radiative and the dielectronic recombination (RR and DR) processes. 
Recombination into group B levels with $\nu > \nu_o$ is
considered as entirely due to DR, neglecting the 
non-resonant background in the energy region where RR is usually extremely
small. The theory by Bell and Seaton (1985) is applied (Nahar and
Pradhan 1994) to calculate the
DR collision strengths.  The $\nu_o$ can assume
any value consistent with the method, but is usually taken to be 10. 
Background photoionization cross sections of the high-Rydberg group 
B levels are computed hydrogenically (referred to as the 
``high-n top-up", Nahar 1996), using  procedures developed by Storey 
and Hummer (1992). 

Several atomic effects related to the present calculations are also
discussed in Nahar \etal (2000). In particular, it is pointed out that
for the H-like and the He-like target ions, with strong dipole allowed
$ 2p \longrightarrow 1s$ and 
$ 1s2p \ (^1P^o_1) \longrightarrow 1s^2 \ (^1S_0)$ transitions
respectively, 
autoionizing resonances are radiatively damped to a significant extent 
and this is taken into account in the
calculation of detailed photoionization cross sections.
In a test study on radiation damping, Pradhan and Zhang (1997)
described a numerical fitting procedure to account for the autoionization
vs. radiative decay of resonances. They also
computed the dielectronic satellites   of Fe~XXIV (e~+~Fe~XXV) and
compared those with experimental data. Zhang \etal (1999) further 
reported BPRM calculations for a few individual
resonance complexes (KLL, KLM, KLN, KLO, KLP), and compared with earlier
works (discussed later).

\section{COMPUTATIONS}

Computations of $\sigma_{\rm PI}$ in the relativistic BPRM intermediate 
coupling approximations are carried out using the package of codes 
from the Iron Project (Berrington \etal 1995; Hummer \etal 1993),
extended from the Opacity Project (OP) codes (Berrington et al. 1987,
OP 1995, 1996). 
Radiation damping of resonances up to $n = 10$ are included through 
use of the extended codes STGF and STGBF (Zhang \etal 1999, Nahar 
and Pradhan 1994). The BPRM calculations are carried out for each total 
angular momentum symmetry $J\pi$, corresponding to a set of fine 
structure target levels $J_t$. The target wavefunctions were obtained
from atomic structure calculations using updated version of the 
code SUPERSTRUCTURE (Eissner et al. 1974)

The level-specific recombination cross sections $\sigma_{\rm RC}(i)$,
into level i of the recombined (e~+~ion) system,
are obtained from the {\it partial} photoionization cross sections
$ \sigma_{\rm PI}(i,g)$ of level i into the ground level of the
recombining ion. These detailed (photo)recombination cross sections are
calculated in the energy region
from threshold up to about $\nu = \nu_o
= 10$, where $\nu$ is the effective quantum number relative to the target
level of the recombining ion. Particular care is taken to delineate the
resonances up to $\nu \leq \nu_o$ completely. The electrons in
this energy range generally recombine to a large number of final (e~+~ion)
levels. Recombination cross sections are computed for all
coupled symmetries and levels, and summed to obtain the total $\sigma_{\rm RC}$.

In the higher energy region, $\nu_o < \nu < \infty$,
below each threshold target level, where the
resonances are narrow and dense and the background is negligible, we
compute the detailed and the resonance averaged DR cross sections.
The DR collision strengths in BPRM are obtained using extensions of 
the $R$-matrix asymptotic region code STGF (Nahar \& Pradhan 1994,
Zhang \etal 1999). It is necessary to use extremely fine energy 
mesh in order to delineate the resonance structures belonging to each 
$n$-complex. 

The level specific recombination rate coefficients are obtained
using a new computer program, BPRRC (Nahar et al 2000). 
The level
specific rates are obtained for energies going up to infinity. 
These rates include both non-resonant and resonant contributions up to
energies $z^2/\nu_o^2$; Contributions from all autoionizing
resonances up to $\nu \leq \nu_o \approx 10$ are included.

The program BPRRC sums up the level specific rates, which is added to
the contributions from the resonant high-n DR, from resonances with
 $\nu_o < \nu < \infty$, to obtain total recombination
rates. As an additional check on the numerical calculations, 
the total recombination rate coefficients, $\alpha_R$, are also 
calculated from the total recombination collision strength,
$\Omega_{RC}$, obtained from all the photoionization cross sections,
and the DR collision strengths. The agreement between the two numerical
approaches is within a few percent.

Finally, the background (non-resonant) contribution from
the high-n states, ($10 < n \leq \infty$), to the total recombination,
is also included as the "top-up" part and is computed in the hydrogenic
approximation (Nahar 1996). This contribution is 
important at low temperatures, but negligible at high temperatures.
The rapid rise in $\alpha_R$ toward
very low temperatures is due to low energy recombination
into the infinite number of these high-n states, at electron energies
not usually high enough for resonant excitations and DR stabilization.

 The program BPRRC is also used to extend the {\it total} photoionization
cross sections in the high energy region, beyond the highest target
threshold in the close coupling wavefunction expansion of the ion, 
by a tail from
Kramers fit of $\sigma_{PI}(E) = \sigma_{PI}^o(E^o/E)^3$, where $E^o$
is the last tabulated energy above all target thresholds.

Below we describe the calculations individually for the ions
under consideration.

\subsection{e + Fe XXV $\longrightarrow$ Fe XXIV}

The fine structure levels of the target ion, Fe~XXV, included in the 
wavefunction expansion for Fe~XXIV are given in Table 1. 
The 13 fine structure levels of Fe~XXV up to $1s3p$ correspond to
configurations $1s^2$, $1s2s$, $1s2p$, $1s3s$, and $1s3p$ (correlation
configurations include those with the $3d$ orbital).  Although
calculated energies are close to one percent of the observed ones, the
latter are used in the computations to obtain accurate positions of 
the resonances. 

All levels of total angular momentum symmetries $1/2 \leq J \leq 11/2$ are 
considered. With largest partial wave of the outer electron $l=7$, 
these correspond to $0\leq L \leq 7$ in doublet and quartet spin 
symmetries. The R-matrix basis set is represented by 30 continuum 
functions. It is necessary to represent the wavefunction expansion 
in the inner region of the R-matrix boundary with a relatively large 
number of terms in order to avoid numerical problems.

\subsection{e + Fe XXVI $\longrightarrow$ Fe XXV}

The wavefunction expansion of Fe XXV is represented by 9 fine structure
levels (Table 1) of hydrogenic Fe XXVI from $1s$ to $3d$. 

The highest partial wave considered is $l=9$ giving $SL\pi$ symmetries 
consisting of $0 \leq L \leq 9$ of singlet and triplet spin symmetries, 
for even and odd parities. All levels of Fe~XXV with total angular 
momentum symmetry $0 \leq J \leq 7$ for even and 8 for odd parity are 
included.  The R-matrix basis set consists of 20 terms.

\section{RESULTS AND DISCUSSION}

Results for photoionization and recombination are presented below, followed
by a discussion of the physical features and effects.

\subsection{Photoionization}

Total and partial ground state cross sections (into the ground and 
excited levels of the residual ion, and into the ground level only,
respectively) are needed for various astrophysical models, such
as in determination of ionization fractions in photoionization
equilibrium, and non-LTE spectral models. Figs. 1 and 2 present the 
ground state photoionization cross section for Fe~XXIV 
($1s^2 \ 2s \ ^2S_{1/2}$) and Fe~XXV $(1s^2 \ ^1S_0)$. Plots (a,b)
in each figure show the partial cross section into the ground level 
(a), and the total cross section (sum of ionization into various 
target levels) (b) of the residual ion. 
For Fe~XXIV and Fe~XXV the first excited target n = 2 threshold(s) 
lie at high energies and the cross sections show  a monotonic decrease 
over a relatively large energy range. The total and the partial cross
sections are identical below the first excited level of the residual
ion, as shown in the figures.
The resonances at high energies belong to Rydberg series of 
$n=2,3$ levels. Owing to the high ion charge z, the resonance complexes 
- groups of levels with the same principal quantum number(s)  - 
are clearly separated, approximately as z$^2$ (discussed in the next section).

The total cross sections Fig. 1(b) show the K-shell ionization jump at 
the n = 2 target levels, i.e. inner-shell photoionization: 

$$ h\nu + Fe~XXIV (1s^2 \ 2s) \longrightarrow e \ + Fe~XXV (1s2s ,1s2p) \ . $$ 

Figure 1(c) displays an expanded view of the resonance structures and 
the inner-shell ionization energy regions. In X-ray photoionization 
models the inner-shell edges play an important role in the overall 
ionization rates.

 In order to fully delineate the resonances for recombination
calculations it is necessary to compute the {\it partial} cross sections at a
very large number of points, typically tens of thousands of energies.
However, for ionization balance calculations it may not be 
necessary to compute the {\it total} photoionization cross sections at an
equally fine mesh since the photoionization rate usually depends on the
convolution over a slowly varying radiation field with frequency.

\subsection{Recombination cross sections and rate coefficients}

Figs. 3(a,b) present the total recombination cross sections
$\sigma_{RC}$,  summed
over all contributing fine structure levels up to n = 10,
for Fe~XXIV and Fe~XXV.  The resonance complexes for 
Fe~XXIV are marked as KLL, KLM, KLN etc., and those for Fe~XXV as LL, 
LM, LN etc. These are the complexes of di-electronic satellite lines 
observed in tokamaks, Electron-Beam-Ion-Traps (EBIT), ion storage 
rings and astrophysical sources. In particular, the KLL complex
has been well studied theoretically (e.g. Gabriel 1972, Bely-Dubau 
\etal 1982), and experimentally (e.g. Beirsdorfer \etal 1992). In an 
earlier work on radiation damping of resonances (Pradhan and Zhang 1997), 
the theoretical intensities of the individual KLL di-electronic 
satellite lines were calculated  and compared with several other 
theoretical calculations and the EBIT experiment (Beirsdorfer \etal 1992). 
The agreement of the KLL satellite lines with experiment is generally 
about 10-20\% (Pradhan and Zhang 1997). Later calculations by Zhang 
\etal (1999) were extended to the KLM, KLN, KLO, and KLP complexes 
and compared with other theoretical calculations (Badnell \etal 1998) 
to investigate the effects of radiation damping on the contribution
of these complexes to the total recombination rate coefficients of
Fe~XXIV.
 
The unified total BPRM recombination rate coefficients, $\alpha_R(T)$,
of $e~+~Fe~XXV \rightarrow Fe~XXIV$ and $e~+~Fe~XXVI \rightarrow Fe~XXV$
averaged over a Maxwellian distribution are 
presented in Table 2. The features of the total recombination rates 
(solid curves) are shown in Fig. 4 for Fe~XXIV, and Fig. 5 for Fe~XXV.
The general features are similar to other ions, as explained in the 
first paper of this series (Nahar \& Pradhan 1997). As pointed out in
our earlier works on unified recombination rates, the total (e~+~ion)
recombination 
rate generally decreases with temperature, dominated essentially by RR
but possibly affected by a low-temperature DR "bump", until the
relatively higher energy resonances enter via DR resulting in a large DR
bump. In Figs. 4 and 5 these 
 high temperatures bumps are at  about
$log_{10}T(K)$ =7.5 for Fe XXIV, and $log_{10}T(K)$ = 7.7 for Fe XXV,
where the total rate rises due to dominance of DR over RR. 
The total recombination
rate coefficients for the hydrogenic Fe XXVI are also given in Table
2 for completeness.

Fig. 4 compares the present total unified BPRM recombination rate 
coefficients (solid curve) for $e~+~Fe~XXV \rightarrow Fe~XXIV$ with 
several other available sets of data. However, since previous works 
treat RR and DR as independent processes, and obtain those
rates individually, we compare the present unified rates with the sum
of the two (RR + DR) rates from earlier works (with the exception of
cases where only one set of data is available).  The dotted curve is 
the sum of the RR and DR rates fitted by Woods \etal (1981), and the
long dashed curves are similar fits by Arnaud and Raymond (1992);
original references to the RR and DR data are given in these references.
The present rates and those by Arnaud and Raymond (1992) agree 
quite well with each other except at very high temperature
where DR dominates. The rates by Arnaud and Raymond show a lower DR peak
than the present one. The rates by Woods \etal (1981) are much higher at
all temperatures than the present rates, or those by Arnaud and Raymond
who adopted the DR rates computed by Karim and Bhalla (1988) 
for individual resonances. 
Woods \etal obtained the RR rates using the Reilman and 
Manson (1978) sub-shell photoionization cross sections in the 
central-field approximation (without LS multiplet structure), and DR 
rates from the Burgess formula (1965). While the present unified rates 
include both the non-resonant (RR) and the resonant recombination (DR), 
previous calculations employ different approximations for these two 
processes from different sources. Therefore it is not possible
to ascertain precisely the causes of these discrepancies.
The short dashed curves are the RR-only rates by Verner and Ferland
(1996), that agree well with present ones in the region (T $< 10^7$
K) where DR is small. It may be noted that the partial rates given in
Zhang et al. (1999) were obtained only for the resonant (i.e. DR)
recombination cross sections, contributions from the low energy 
background cross sections were not included.

Total $\alpha_R(T)$ for $e + Fe~XXVI \rightarrow Fe~XXV$ are compared 
with others in Fig. 5. The unified total recombination rates are
compared with the (RR + DR) rates from Woods \etal (1981, dotted), 
Arnaud and Raymond (1992, long-dashed), and Bely-Dubau et al. (1982, 
dot-long-dashed). The unified rates agree well with all except at high
temperatures where the rates by Woods \etal are higher, and those by Arnaud
and Raymond are lower. The DR-only rates 
by Romanik (1988, short-dash-dot) are much lower (without the RR 
contribution). The RR-only rates by Verner and Ferland 
(1996, short-dashed) are in reasonable agreement with the present rates 
up to about $log_{10}T$ = 7.3, in the temperature range with negligible DR 
contributions.

\subsubsection{Level-specific cross sections and rate coefficients}

In Figs. 6 and 7 we show the level-specific recombination 
rate  coefficients into the ground and the excited bound levels of Fe~XXIV,
for the $1s^2 \ ns, \ np $ Rydberg series up to n = 10. These are
the first such calculations, and level-specific data have been obtained
for all $\ell \leq 9$ and associated $J\pi$ symmetries. 
The behavior of the level-specific rates
mimics that of the total (this is only true for the simple ions
under consideration; in general the level specific rates show
significantly different structure for complex ions, as seen in our
previous works on low ionization stages of iron, Fe~I~--~V, for example).
The only distinguishing feature is the slight DR bump. Since
the computations are enormously involved, particularly related
to the resolution of resonances, 
the consistency of the individual level-specific rates along the Rydberg
series is an indication of numerical precision.

Fig. 8 presents level specific recombination rate coefficients of $1sns
(^3S)$ Rydberg series of Fe~XXV levels up to $n$ = 10. The features are
similar to those of Fe~XXIV. Although resolution
of resonances in each cross section is very cumbersome, the sum of the
level-specific rate coefficients, together with the DR contribution,
agrees within a few percent with the total recombination rate
coefficient obtained from the total collision strengths, 
thus providing a numerical and self-consistency check.

\subsubsection{X-ray transitions w,x,y,z in Fe~XXV -- photoionization 
and recombination}

The first detailed BPRM calculations for level-specific photoionization 
from, and recombination into, the n = 2 levels of Fe~XXV are presented.
Figs. 9 and 10 respectively show the level-specific photoionization cross 
sections, and  recombination rate coefficients, for the ground and
the excited n  = 2 levels that are of considerable importance in X-ray
spectroscopy as they are responsible for the formation of the
w,x,y,z lines from the 4 transitions $1s^2 \ (^1S_0) 
\longleftarrow 1s2p (^1P^o_1), 1s2p (^3P^o_2), 1s2p (^3P^o_1), 1s2s
(^3S_1)$, respectively. The present work is particularly relevant to the 
formation of these X-ray lines since recombination-cascades from excited 
levels play an important role in determining the intensity ratios in 
coronal equilibrium and non-equilibrium plasmas (Pradhan 1985). 
The cross sections in Fig. 9 show the K-shell ionization jump at
the n = 2 target levels, i.e., from photoionization as

$$ h\nu + Fe~XXV (1s2s, 1s2p) \longrightarrow e \ + Fe~XXVI (2s \ ,2p) \ . $$

The level-specific rates in Fig. 10 are in reasonable agreement with 
those obtained by Mewe \& Schrijver (1978, hereafter MS), that have 
been widely employed in the calculation of X-ray spectra of 
He-like ions (e.g. Pradhan 1982). This is quite unlike the case for 
He-like C~V where the unified level-specific rates differed 
considerably with those of MS (Nahar \etal 2000).  We compare with
the direct (RR + DR) rates, separately calculated by MS using 
approximate Z-scaled RR and DR rates for the individual n = 2 levels 
of He-like ions. Their RR rates were from Z-scaled recombination rate 
of He II given by Burgess \& Seaton (1960); the LS coupling data were 
divided according to the statistical weights of the fine structure 
levels. The MS DR rates were obtained using averaged autoionization 
probabilities for iron calculated with hydrogenic
wavefunctions, together with radiative decay probabilities of
the resonant $2s2p, 2p^2, (2p \ 3s, \ 2p3p, \ 2p3d)$ levels, 
decaying to the final n = 2 levels. Although the present 
work includes DR contributions from all resonances 
up to $2p \ n \ell;  \ n \leq 10, \ell \leq n-1 $ (Figs. 3a,b), the 
final values appear to agree well.
The most plausible explanation for the good agreement with MS is that
their rates were optimized especially for Fe~XXV.

Using the present level-specific data, recombination-cascade matrices 
may now be constructed for Fe~XXIV and Fe~XXV,  to obtain
effective recombination rates into specific fine structure levels nSLJ,
with $n \leq 10$ and $\ell \leq n-1 $ (e.g. Pradhan 1985). 
The present data is more than sufficient
for extrapolation to high-n,$\ell$ necessary to account for 
all cascade contributions. Also needed are the radiative transition 
probabilities for all fine structure levels of Fe~XXIV and Fe~XXV, 
up to the n = 10 levels; those have also been calculated using the
BPRM method under the Iron Project  (Nahar and Pradhan 1999).

A discussion of some of the important atomic effects 
such as resolution and radiation damping of resonances, interference
between non-resonant (RR) and resonant (DR) recombination, comparison
with experimental data and uncertainties, and general features of the
unified (e~+~ion) recombination rates,  has been given in the first
paper on the new BPRM calculations -- paper IV in the persent series 
(Nahar \etal 2000).

\section{CONCLUSION}

 New relativistic calculations for total and level-specific
photoionization and recombination are presented for 
Fe~XXIV and Fe~XXV of general interest in X-ray spectroscopy of
laboratory and astrophysical sources.
The di-electronic satellite rates for the KLL complex of
$e~+~Fe~XXV \rightarrow Fe~XXIV$, and for several higher complexes,
have earlier been shown to be in very good agreement with experiments
and other theoretical calculations (Pradhan and Zhang 1997; Zhang \etal 1999), 
to about 10-20\%;
it is therefore expected that the present rates should be
definitive, with similar uncertainty.

The unified theoretical formulation
and experimental measurements both
suggest that the unphysical and imprecise division of the
recombination process into `radiative recombination (RR)' and
`di-electronic recombination (DR)' be replaced by
`non-resonant' and `resonant' recombination, since these are naturally
inseparable.

 Further calculations are in progress for Oxygen (O~VI and O~VII).

The available data includes: 

(A) Photoionization cross sections for bound fine structure levels
of Fe~XXIV and Fe~XXV up to the n = 10 complexes -- both the total and 
the partial (into the ground
level of the residual ion).

(B) Total, unified recombination rates for Fe~XXIV and Fe~XXV, and 
level-specific recombination rate coefficients for levels up to n = 10. 

(C) Recombination rate coefficients for H-like Fe~XXVI, computed in LS
coupling and included for completeness 
for the computation of ionization fractions towards the high ionization
end.

All photoionization and recombination data are available electronically
from the first author at: nahar@astronomy.ohio-state.edu. The total
recombination rate coefficients are also available from the Ohio State
Atomic Astrophysics website at: www.astronomy.ohio-state.edu/$\sim$pradhan.

%

\acknowledgments

This work was supported partially by grants from NSF (AST-9870089),
NASA (NAG5-8423,EL9-1013A). The computational work was carried out on 
the Cray T94 at the Ohio Supercomputer Center.

\clearpage

\begin{table}
\caption{Target terms in the eigenfunction expansions of Fe XXV and Fe
XXVI. The target energies are in Rydbergs.
}
\scriptsize
\begin{tabular}{llll}
\hline
\multicolumn{2}{c}{Fe XXV} & \multicolumn{2}{c}{Fe XXVI} \\
\hline
1s$^2(^1{\rm S}_0)$     & 0.0    &
1s$(^2{\rm S}_{1/2})$   &   0.00 \\
1s2s$(^3{\rm S}_1)$     & 487.774760 &
2p$(^2{\rm P}^o_{1/2})$   & 510.9598  \\
1s2p$(^3{\rm P^o}_0)$     & 489.899743 &
2s$(^2{\rm S}_{1/2})$ & 511.0012  \\
1s2p$(^3{\rm P}^o_1)$   & 490.071608 &
2p$(^2{\rm P}\o_{3/2})$ & 512.5190 \\
1s2s$(^1{\rm S}_0)$   & 490.091292 &
3p$(^2{\rm P}^o_{1/2})$   & 606.0989 \\
1s2p$(^3{\rm P}^o_2)$   & 491.132414 &
3s$(^2{\rm S}_{1/2})$ & 606.1118 \\
1s2p$(^1{\rm P}^o_1)$   & 492.448740 &
3p$(^2{\rm P}^o_{3/2})$ & 606.5603 \\
1s3s$(^3{\rm S}_1)$     & 579.251214 &
3d$(^2{\rm D}_{5/2})$   & 606.5612 \\
1s3p$(^3{\rm P}_0)$     & 579.251214 &
3d$(^2{\rm D}_{3/2})$   & 606.71180 \\
1s3s$(^1{\rm S}_0)$   & 579.251214 & & \\
1s3p$(^3{\rm P}^o_1)$   & 579.251214 & & \\
1s3p$(^3{\rm P}^o_2)$   & 579.251214 & & \\
1s3p$(^1{\rm P}^o_1)$   & 579.251214 & & \\
\multicolumn{2}{c}{13-CC}  & \multicolumn{2}{c}{9-CC} \\
\hline
\end{tabular}
\end{table}

\pagebreak

\begin{table}
\caption{Total recombination rate coefficients, $\alpha_R(T)$, of Fe XXIV,
Fe XXV and Fe XXVI. }
\scriptsize
\begin{tabular}{crrrcrrr}
\hline
$log_{10}T$ & \multicolumn{3}{c}{$\alpha_R(cm^3s^{-1})$} &
$log_{10}T$ & \multicolumn{3}{c}{$\alpha_R(cm^3s^{-1})$}\\
(K) & \multicolumn{1}{c}{Fe XXIV} & \multicolumn{1}{c}{Fe XXV} &
\multicolumn{1}{c}{Fe XXVI} & (K) & \multicolumn{1}{c}{Fe XXIV} & 
\multicolumn{1}{c}{Fe XXV} & \multicolumn{1}{c}{Fe XXVI} \\
\hline
  1.0 &  2.38E-08 & 2.79E-08 & 3.21E-08 &
  5.1 &  9.67E-11 & 1.24E-10 & 1.52E-10 \\
  1.1 &  2.11E-08 & 2.48E-08 & 2.85E-08 &
  5.2 &  8.30E-11 & 1.07E-10 & 1.32E-10 \\
  1.2 &  1.88E-08 & 2.21E-08 & 2.54E-08 &
  5.3 &  7.12E-11 & 9.18E-11 & 1.14E-10 \\
  1.3 &  1.67E-08 & 1.96E-08 & 2.25E-08 &
  5.4 &  6.10E-11 & 7.89E-11 & 9.85E-11 \\
  1.4 &  1.48E-08 & 1.74E-08 & 2.00E-08 &
  5.5 &  5.22E-11 & 6.77E-11 & 8.51E-11 \\
  1.5 &  1.31E-08 & 1.55E-08 & 1.78E-08 &
  5.6 &  4.47E-11 & 5.80E-11 & 7.34E-11 \\
  1.6 &  1.16E-08 & 1.37E-08 & 1.58E-08 &
  5.7 &  3.81E-11 & 4.96E-11 & 6.33E-11 \\
  1.7 &  1.03E-08 & 1.22E-08 & 1.40E-08 &
  5.8 &  3.25E-11 & 4.23E-11 & 5.45E-11 \\
  1.8 &  9.13E-09 & 1.08E-08 & 1.24E-08 &
  5.9 &  2.77E-11 & 3.61E-11 & 4.69E-11 \\
  1.9 &  8.06E-09 & 9.53E-09 & 1.10E-08 &
  6.0 &  2.35E-11 & 3.07E-11 & 4.02E-11 \\
  2.0 &  7.12E-09 & 8.42E-09 & 9.71E-09 &
  6.1 &  1.99E-11 & 2.61E-11 & 3.46E-11 \\
  2.1 &  6.28E-09 & 7.44E-09 & 8.58E-09 &
  6.2 &  1.69E-11 & 2.22E-11 & 2.96E-11 \\
  2.2 &  5.53E-09 & 6.56E-09 & 7.58E-09 &
  6.3 &  1.42E-11 & 1.88E-11 & 2.53E-11 \\
  2.3 &  4.87E-09 & 5.78E-09 & 6.69E-09 &
  6.4 &  1.20E-11 & 1.59E-11 & 2.17E-11 \\
  2.4 &  4.29E-09 & 5.09E-09 & 5.90E-09 &
  6.5 &  1.00E-11 & 1.34E-11 & 1.85E-11 \\
  2.5 &  3.76E-09 & 4.48E-09 & 5.20E-09 &
  6.6 &  8.40E-12 & 1.13E-11 & 1.58E-11 \\
  2.6 &  3.30E-09 & 3.94E-09 & 4.57E-09 &
  6.7 &  7.00E-12 & 9.52E-12 & 1.34E-11 \\
  2.7 &  2.90E-09 & 3.46E-09 & 4.02E-09 &
  6.8 &  5.82E-12 & 8.01E-12 & 1.14E-11 \\
  2.8 &  2.54E-09 & 3.04E-09 & 3.53E-09 &
  6.9 &  4.83E-12 & 6.72E-12 & 9.64E-12 \\
  2.9 &  2.22E-09 & 2.67E-09 & 3.10E-09 &
  7.0 &  4.02E-12 & 5.65E-12 & 8.14E-12 \\
  3.0 &  1.94E-09 & 2.33E-09 & 2.72E-09 &
  7.1 &  3.41E-12 & 4.80E-12 & 6.88E-12 \\
  3.1 &  1.70E-09 & 2.04E-09 & 2.39E-09 &
  7.2 &  2.98E-12 & 4.13E-12 & 5.78E-12 \\
  3.2 &  1.48E-09 & 1.79E-09 & 2.09E-09 &
  7.3 &  2.69E-12 & 3.62E-12 & 4.85E-12 \\
  3.3 &  1.29E-09 & 1.56E-09 & 1.83E-09 &
  7.4 &  2.50E-12 & 3.23E-12 & 4.06E-12 \\
  3.4 &  1.13E-09 & 1.36E-09 & 1.60E-09 &
  7.5 &  2.33E-12 & 2.89E-12 & 3.38E-12 \\
  3.5 &  9.81E-10 & 1.19E-09 & 1.40E-09 &
  7.6 &  2.14E-12 & 2.58E-12 & 2.81E-12 \\
  3.6 &  8.54E-10 & 1.04E-09 & 1.23E-09 &
  7.7 &  1.92E-12 & 2.27E-12 & 2.32E-12 \\
  3.7 &  7.42E-10 & 9.07E-10 & 1.07E-09 &
  7.8 &  1.67E-12 & 1.95E-12 & 1.92E-12 \\
  3.8 &  6.45E-10 & 7.90E-10 & 9.35E-10 &
  7.9 &  1.41E-12 & 1.64E-12 & 1.57E-12 \\
  3.9 &  5.60E-10 & 6.89E-10 & 8.16E-10 &
  8.0 &  1.16E-12 & 1.36E-12 & 1.28E-12 \\
  4.0 &  4.86E-10 & 5.99E-10 & 7.11E-10 &
  8.1 &  9.33E-13 & 1.10E-12 & 1.04E-12 \\
  4.1 &  4.22E-10 & 5.22E-10 & 6.21E-10 &
  8.2 &  7.35E-13 & 8.80E-13 & 8.40E-13 \\
  4.2 &  3.65E-10 & 4.53E-10 & 5.41E-10 &
  8.3 &  5.69E-13 & 6.94E-13 & 6.73E-13 \\
  4.3 &  3.16E-10 & 3.93E-10 & 4.70E-10 &
  8.4 &  4.35E-13 & 5.41E-13 & 5.37E-13 \\
  4.4 &  2.73E-10 & 3.42E-10 & 4.10E-10 &
  8.5 &  3.29E-13 & 4.19E-13 & 4.26E-13 \\
  4.5 &  2.36E-10 & 2.96E-10 & 3.56E-10 &
  8.6 &  2.47E-13 & 3.22E-13 & 3.36E-13 \\
  4.6 &  2.04E-10 & 2.57E-10 & 3.10E-10 &
  8.7 &  1.84E-13 & 2.46E-13 & 2.63E-13 \\
  4.7 &  1.76E-10 & 2.23E-10 & 2.69E-10 &
  8.8 &  1.36E-13 & 1.87E-13 & 2.05E-13 \\
  4.8 &  1.52E-10 & 1.92E-10 & 2.34E-10 &
  8.9 &  1.00E-13 & 1.41E-13 & 1.59E-13 \\
  4.9 &  1.31E-10 & 1.67E-10 & 2.03E-10 &
  9.0 &  7.36E-14 & 1.06E-13 & 1.22E-13 \\
  5.0 &  1.12E-10 & 1.44E-10 & 1.76E-10 &
  & & & \\
\hline
\end{tabular}
\end{table}


\clearpage

%
%

\def\amp{{Adv. At. Molec. Phys.}\ }
\def\apj{{ Astrophys. J.}\ }
\def\apjs{{Astrophys. J. Suppl. Ser.}\ }
\def\apjl{{Astrophys. J. (Letters)}\ }
\def\aj{{Astron. J.}\ }
\def\aa{{Astron. Astrophys.}\ }
\def\aasup{{Astron. Astrophys. Suppl.}\ }
\def\adndt{{At. Data Nucl. Data Tables}\ }
\def\cpc{{Comput. Phys. Commun.}\ }
\def\jqsrt{{J. Quant. Spect. Radiat. Transfer}\ }
\def\jpb{{Journal Of Physics B}\ }
\def\pasp{{Pub. Astron. Soc. Pacific}\ }
\def\mn{{Mon. Not. R. astr. Soc.}\ }
\def\pra{{Physical Review A}\ }
\def\prl{{Physical Review Letters}\ }
\def\zpds{{Z. Phys. D Suppl.}\ }
\def\adndt{Atomic Data And Nuclear Data Tables}

%

\begin{figure}
\centering
\psfig{figure=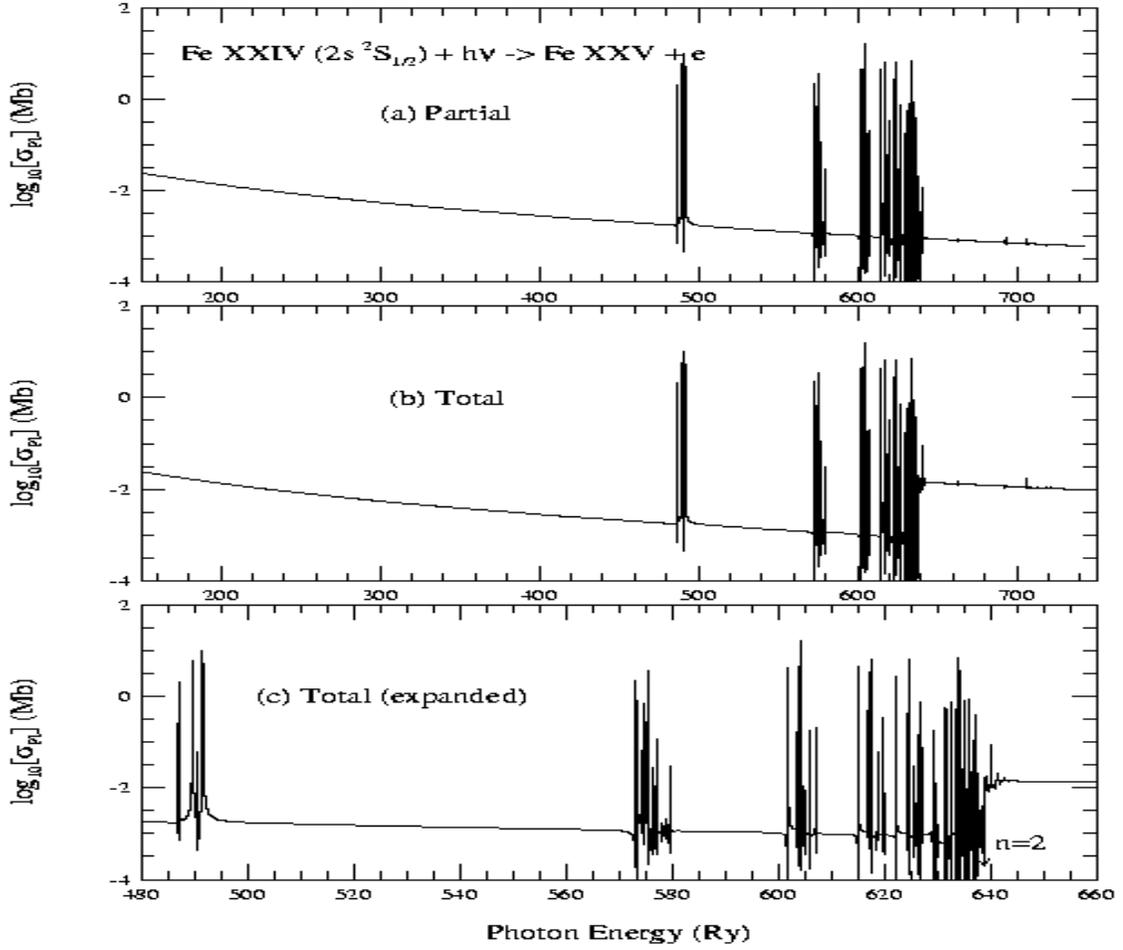,height=15.0cm,width=18.0cm}
\caption{Photoionization of the ground state $1s^2 \ 2s \ (^2S_{1/2})$ 
of Fe~XXIV: (a) partial cross section into the ground level 
$1s^2 \ (^1S_0)$ of Fe~XXV ; (b) total cross section;
(c) an expanded view of the resonances and inner-shell thresholds.
The large jump in (b) corresponds to the K-shell ionization edge.}
\end{figure}

\begin{figure}
\centering
\psfig{figure=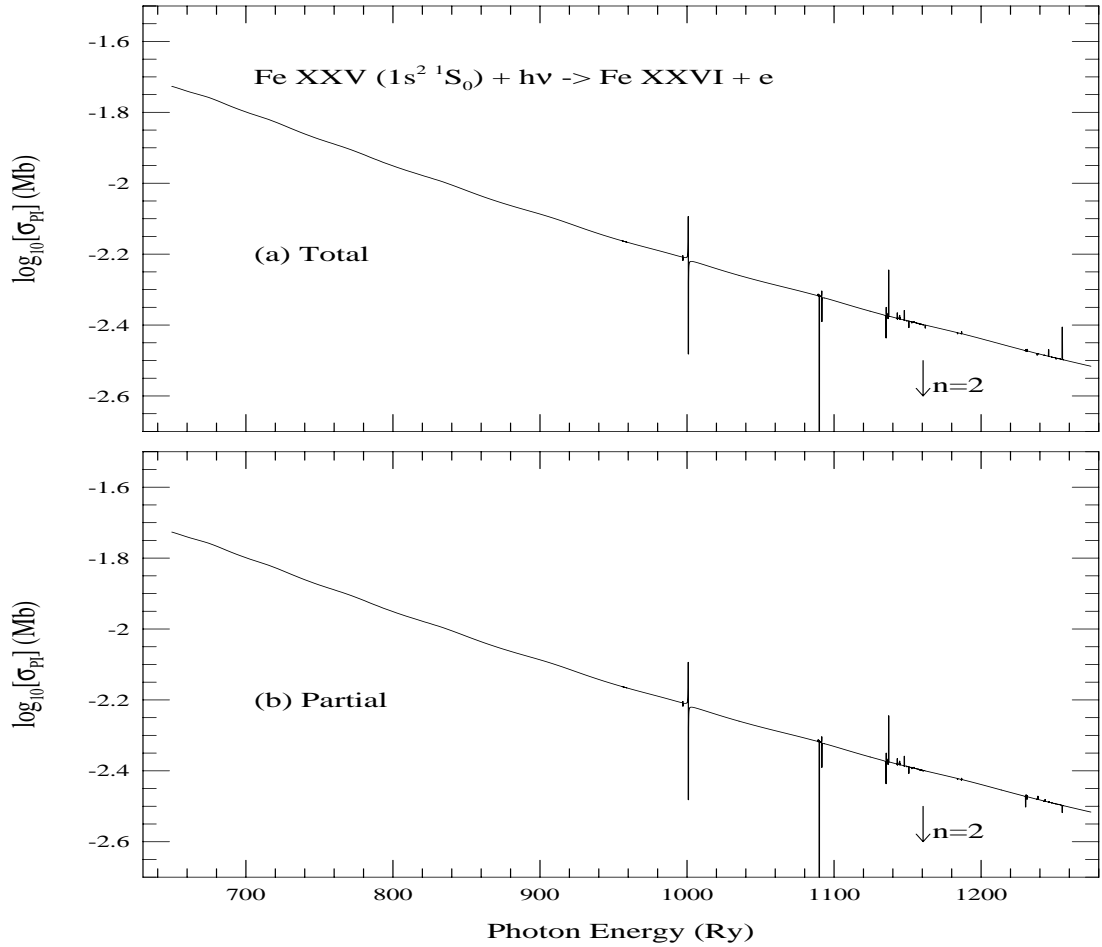,height=15.0cm,width=18.0cm}
\caption{Photoionization of the ground state $1s^2 \ (^1S_0)$ of Fe~XXV: 
(a) partial cross section into the ground level 
$1s \ (^2S_{1/2})$ of Fe~XXVI ; (b) total cross section.}
\end{figure}

\begin{figure}
\centering
\psfig{figure=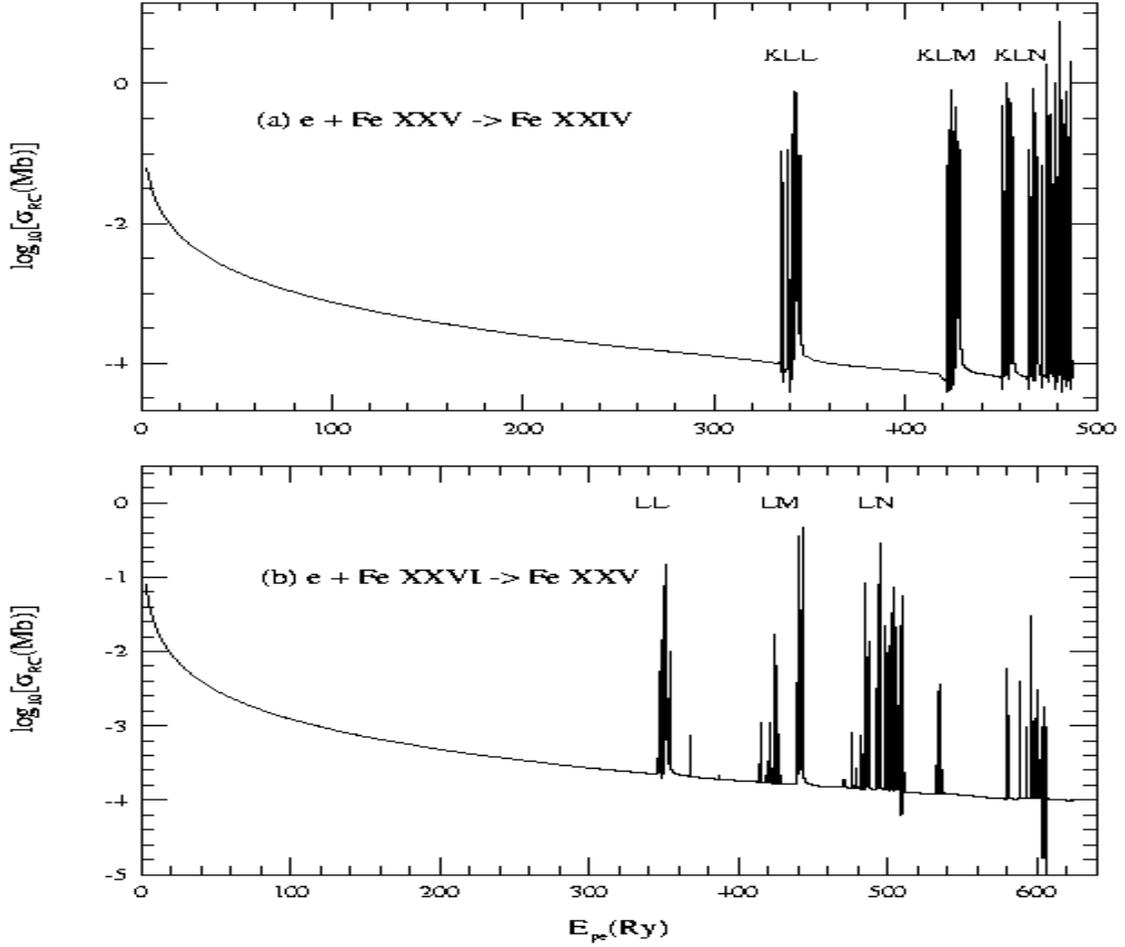,height=15.0cm,width=18.0cm}
\caption{Total unified (e + ion) photo-recombination cross sections, 
$\sigma_{RC}$, of (a) Fe~XXIV and (b) Fe~XXV. Note that the $\sigma_{RC}$
exhibit considerably more resonance structures than the corresponding 
ground level $\sigma_{PI}$ in Figs. 1 and 2, since the former are 
summed over the ground {\it and} many excited recombined levels.}
\end{figure}

\begin{figure}
\centering
\psfig{figure=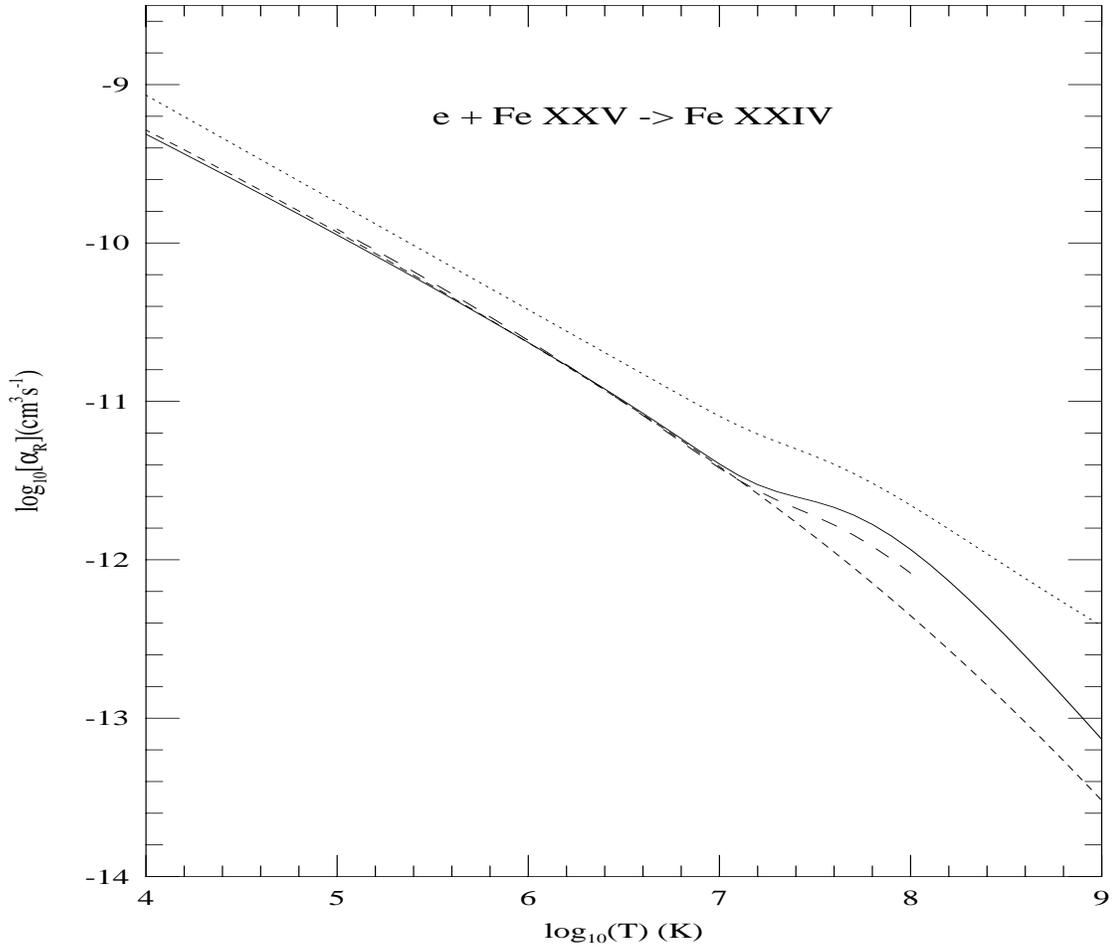,height=15.0cm,width=18.0cm}
\caption{ Total unified rate coefficients for Fe~XXIV:
BPRM with fine structure (solid curve); sum of (RR + DR) rates : from
Arnaud and Raymond 1992 (long-dash); from Woods \etal 1981 (dotted);
RR rates from Verner and Ferland 1996 (short-dash).}
\end{figure}

\begin{figure}
\centering
\psfig{figure=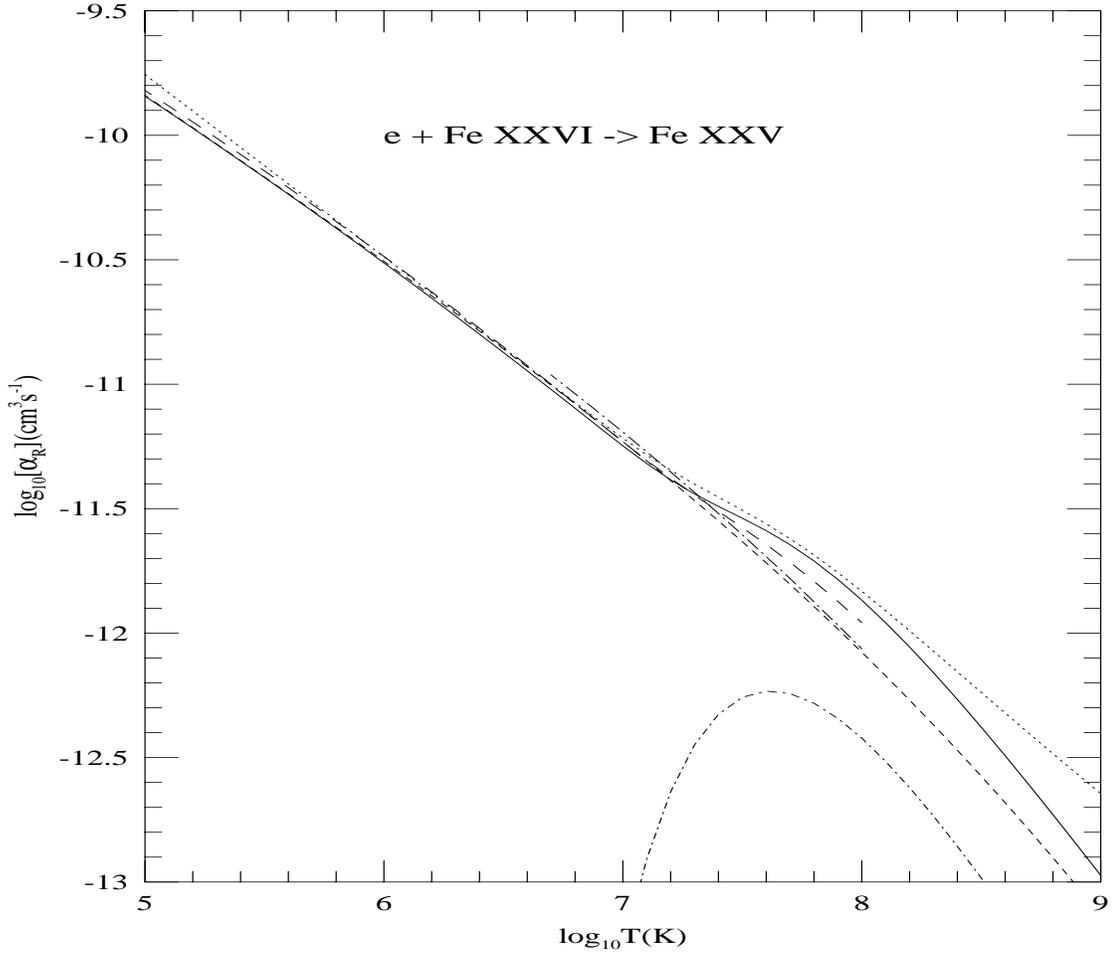,height=15.0cm,width=18.0cm}
\caption{Total unified rate coefficients for Fe~XXV: BPRM with fine 
structure (solid curve); sum of (RR + DR) rates: from Arnaud and Raymond 
1992 (long-dash), from Woods \etal 1981 (dotted), from Bely-Dubau et al.
1982 (dot-long-dash); RR-only rates from Verner and Ferland 1996 
(short-dash); DR-only rates from Romanik 1988 (dot-dash). }
\end{figure}

\begin{figure}
\centering
\psfig{figure=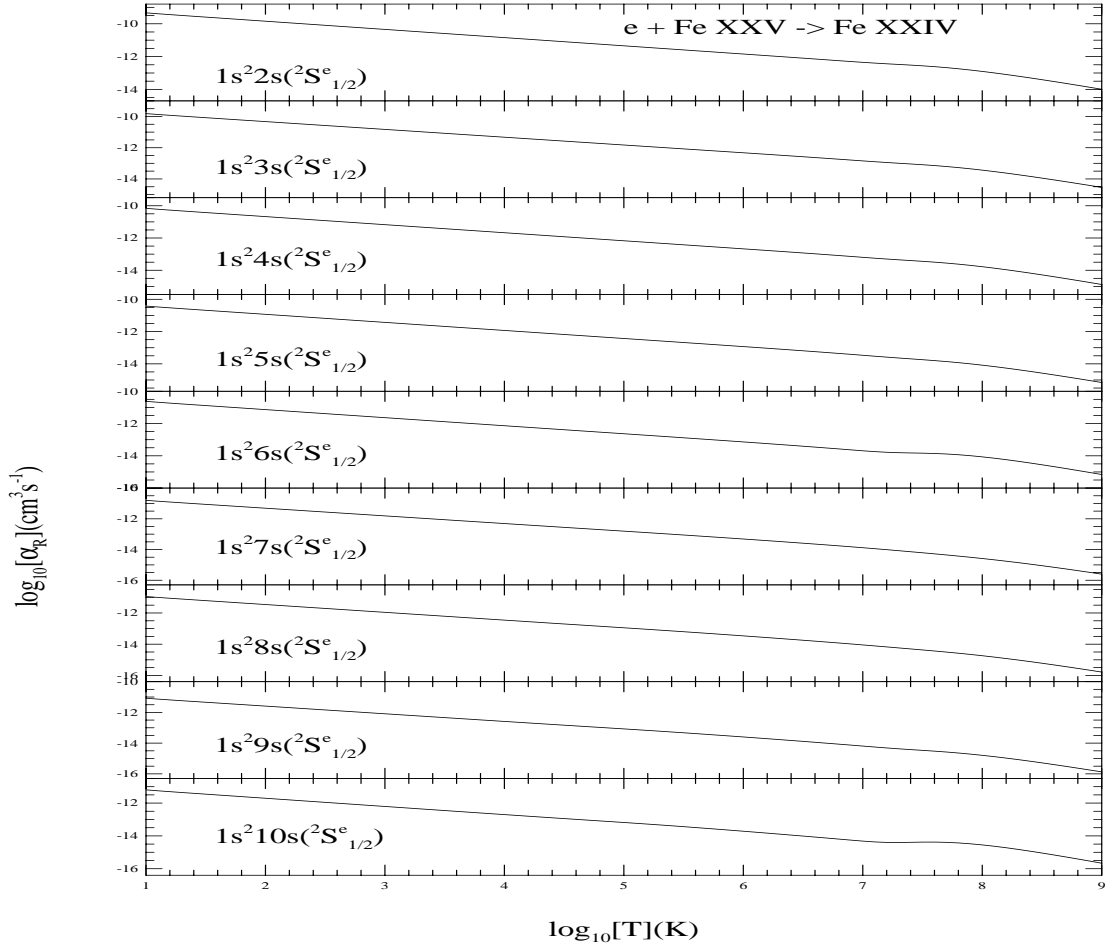,height=15.0cm,width=18.0cm}
\caption{ Level-specific recombination rate coefficients for Fe~XXIV 
into the ground and excited levels of the $1s^2 \ ns (^2S^e_{1/2})$ 
Rydberg series, $n \leq 10$.}
\end{figure}

\begin{figure}
\centering
\psfig{figure=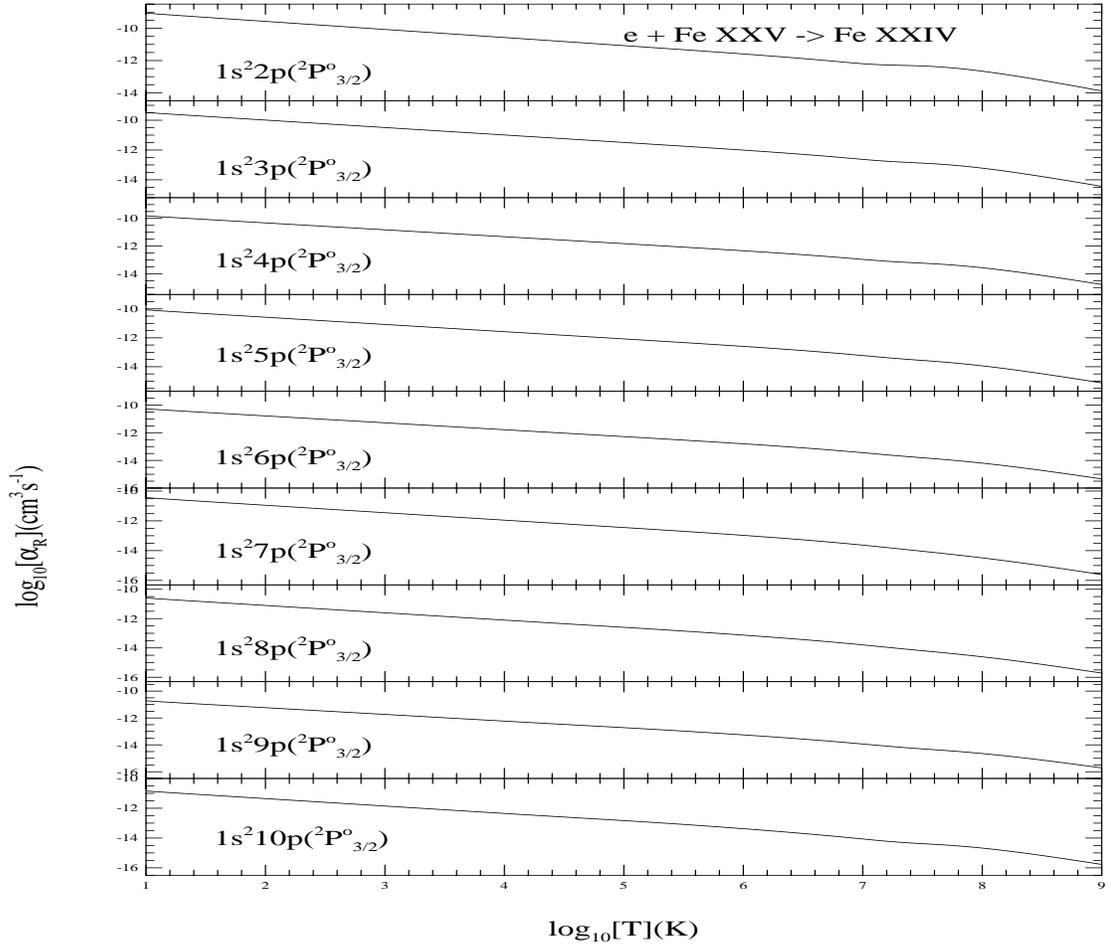,height=15.0cm,width=18.0cm}
\caption{  Level-specific recombination rate coefficients for
Fe~XXIV into the excited levels of
the $1s^2 \ np \ (^2P^o_{3/2})$ Rydberg series, $n \leq 10$.}
\end{figure}

\begin{figure}
\centering
\psfig{figure=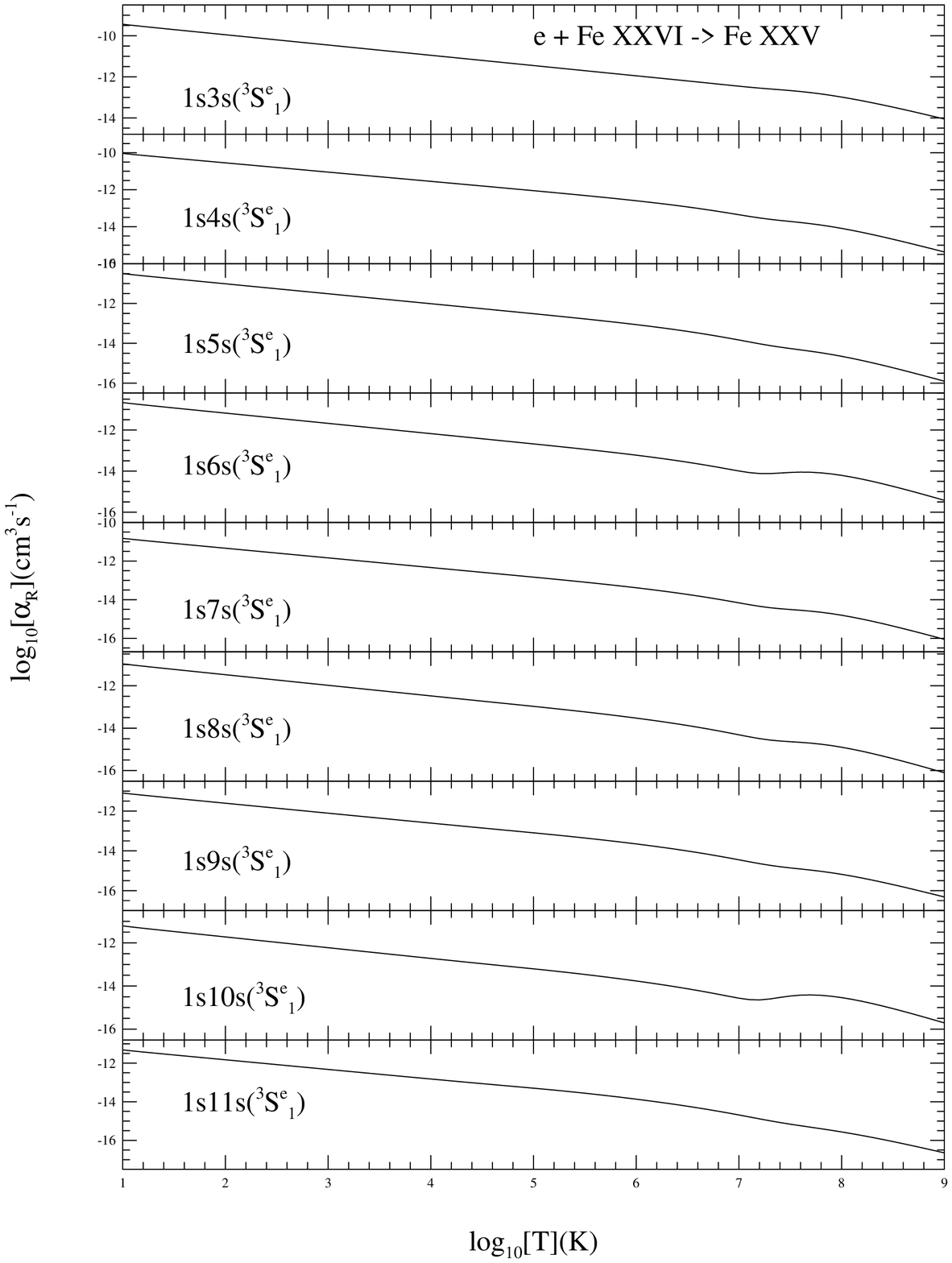,height=15.0cm,width=18.0cm}
\caption{  Level-specific recombination rate coefficients for
Fe~XXV into the excited levels of
the $1s \ ns \ (^3S^e_1)$ Rydberg series, $n \leq 11$.}
\end{figure}

\begin{figure}
\centering
\psfig{figure=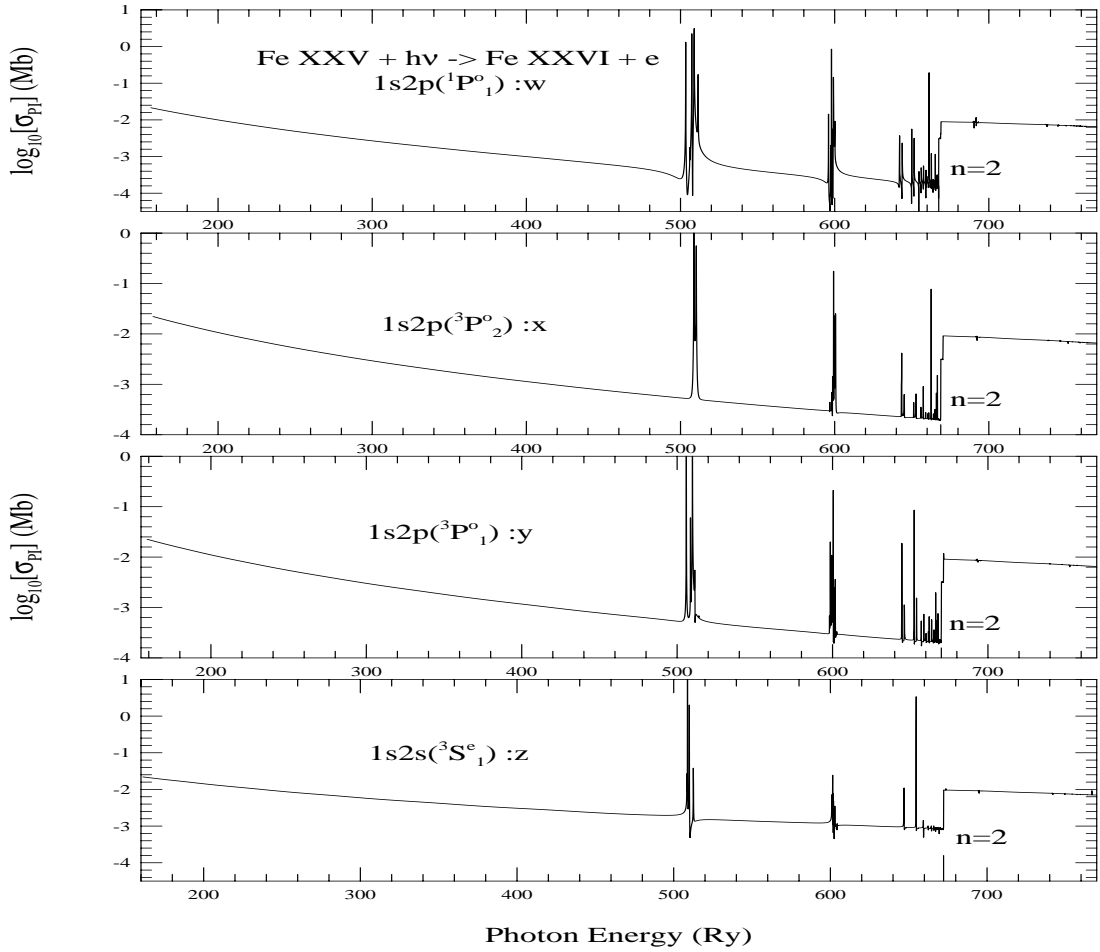,height=15.0cm,width=18.0cm}
\caption{Total photoionization cross sections of the
excited  n = 2 levels of Fe~XXV. The levels shown are the
ones responsible for the prominent X-ray lines w,x,y, and z.}
\end{figure}

\begin{figure}
\centering
\psfig{figure=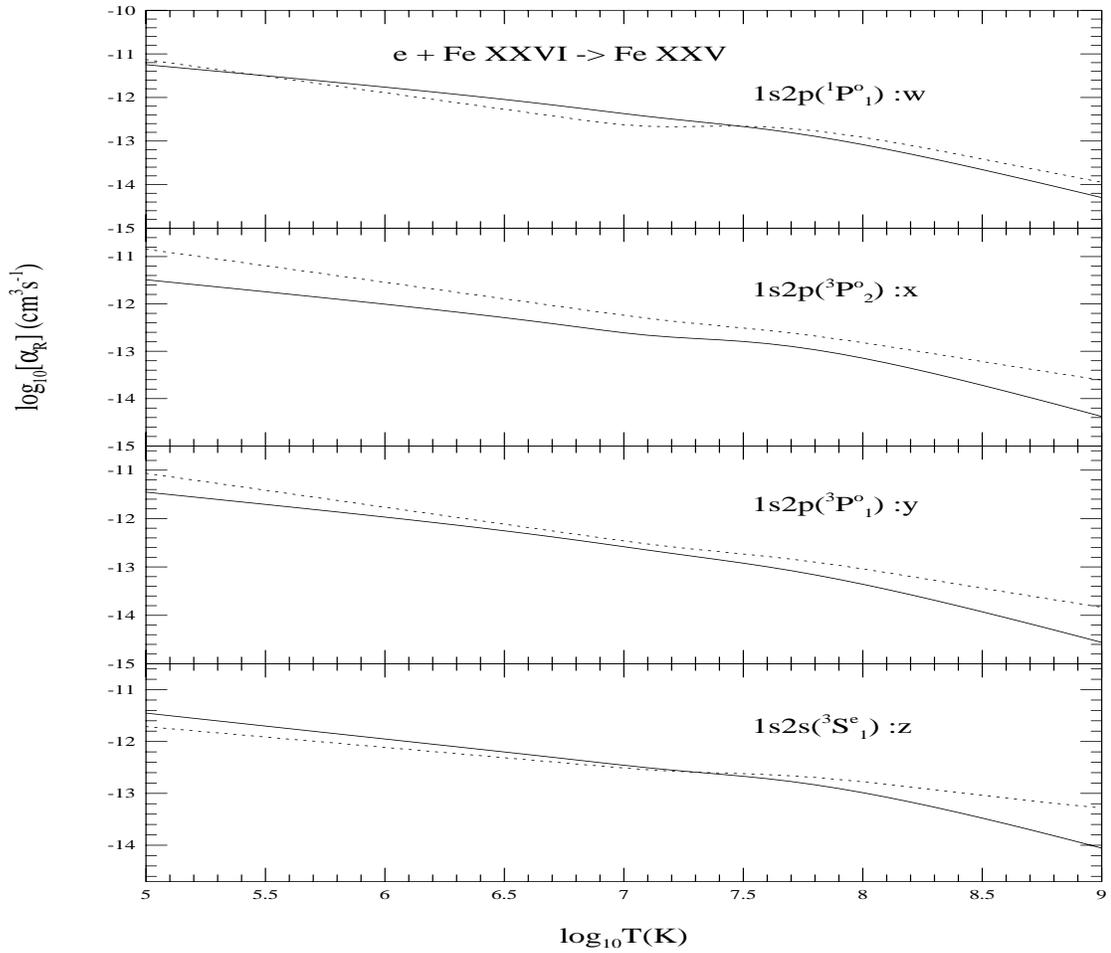,height=15.0cm,width=18.0cm}
\caption{Level-specific recombination rate coefficients for 
Fe~XXV into the excited  n = 2 levels - present (solid); Mewe and 
Schrijver (1978, dotted). The levels shown are the
ones responsible for the prominent X-ray lines w,x,y, and z.}
\end{figure}

\newpage

\end{document}